\documentclass[aps,prc,preprint,superscriptaddress,showpacs,amsmath]{revtex4}
\usepackage{amssymb}
\usepackage{graphicx}

\begin{document}

%\preprint{}

\title{Isovector and isoscalar spin-flip M1 strengths in $^{11}$B}


\author{T.~Kawabata}
\email[]{kawabata@cns.s.u-tokyo.ac.jp}
%\thanks{}
\affiliation{Center for Nuclear Study (CNS), University of Tokyo, RIKEN
campus, 2-1 Hirosawa, Wako, Saitama 351-0198, Japan}

\author{H.~Akimune}
\affiliation{Department of Physics, Konan University,
Kobe, Hyogo 658-8501, Japan}
\author{H.~Fujimura}
\affiliation{Department of Physics, Kyoto University,
Kyoto 606-8502, Japan}
\author{H.~Fujita}
\affiliation{
School of Physics, University of the Witwatersrand,
Johannesburg, 2050, South Africa.}
\author{Y.~Fujita}
\affiliation{Department of Physics, Osaka University,
Toyonaka, Osaka 560-0043, Japan}
\author{M.~Fujiwara}
\affiliation{Research Center for Nuclear Physics, 
Osaka University, Ibaraki, Osaka 567-0047, Japan}
\affiliation{Advanced Photon Research Center, Japan Atomic Energy
Research Institute, Kizu, Kyoto 619-0215, Japan}
\author{K.~Hara}
\affiliation{
KEK, High Energy Accelerator Research Organization, Tsukuba, Ibaraki
305-0801, Japan.}
\author{K.~Y.~Hara}
\affiliation{Department of Physics, Konan University,
Kobe, Hyogo 658-8501, Japan}
\author{K.~Hatanaka}
\affiliation{Research Center for Nuclear Physics, 
Osaka University, Ibaraki, Osaka 567-0047, Japan}
\author{T.~Ishikawa}
\affiliation{
Laboratory of Nuclear Science, Tohoku University, Sendai, 
Miyagi 982-0216, Japan.}
\author{M.~Itoh}
\affiliation{Research Center for Nuclear Physics, 
Osaka University, Ibaraki, Osaka 567-0047, Japan}
\author{J.~Kamiya}
\affiliation{
Japan Atomic Energy Institute, Tokai, Ibaraki 319-1195, Japan.}
\author{S.~Kishi}
\affiliation{Department of Physics, Kyoto University,
Kyoto 606-8502, Japan}
\author{M.~Nakamura}
\affiliation{
Wakayama Medical University, Wakayama 641-8509, Japan.}
\author{K.~Nakanishi}
\affiliation{Research Center for Nuclear Physics, 
Osaka University, Ibaraki, Osaka 567-0047, Japan}
\author{T.~Noro}
\affiliation{Department of Physics, Kyushu University, 
Fukuoka 812-8581, Japan}
\author{H.~Sakaguchi}
\affiliation{Department of Physics, Kyoto University,
Kyoto 606-8502, Japan}
\author{Y.~Shimbara}
\affiliation{Research Center for Nuclear Physics, 
Osaka University, Ibaraki, Osaka 567-0047, Japan}
\author{H.~Takeda}
\affiliation{
RIKEN (The Institute for Physical and Chemical Research),
Wako, Saitama 351-0198, Japan.}
\author{A.~Tamii}
\affiliation{Research Center for Nuclear Physics, 
Osaka University, Ibaraki, Osaka 567-0047, Japan}
\author{S.~Terashima}
\affiliation{Department of Physics, Kyoto University,
Kyoto 606-8502, Japan}
\author{H.~Toyokawa}
\affiliation{Japan Synchrotron Radiation Research Institute, 
Hyogo 679-5198, Japan}
\author{M.~Uchida}
\affiliation{Department of Physics, Kyoto University,
Kyoto 606-8502, Japan}
\author{H.~Ueno}
\affiliation{
RIKEN (The Institute for Physical and Chemical Research),
Wako, Saitama 351-0198, Japan.}
\author{T.~Wakasa}
\affiliation{Department of Physics, Kyushu University, 
Fukuoka 812-8581, Japan}
\author{Y.~Yasuda}
\affiliation{Department of Physics, Kyoto University,
Kyoto 606-8502, Japan}
\author{H.~P.~Yoshida}
\affiliation{Research Center for Nuclear Physics, 
Osaka University, Ibaraki, Osaka 567-0047, Japan}
\author{M.~Yosoi}
\affiliation{Department of Physics, Kyoto University,
Kyoto 606-8502, Japan}

\date{\today}

\begin{abstract}
The $^{11}$B($^3$He$,\,t$), $^{11}$B($d,\,d'$), and $^{11}$B($p,\,p'$)
reactions were measured at forward scattering angles including $0^\circ$
to study the isovector and isoscalar spin-flip M1 strengths in $^{11}$B.
The measured $^{11}$B($^3$He$,\,t$) cross sections were compared with
the results of the distorted-wave impulse-approximation (DWIA)
calculation, and the
Gamow-Teller (GT) strengths for low-lying states in $^{11}$C were
determined. The GT strengths were converted to the isovector spin-flip
M1 strengths using the isobaric analog relations under the assumption of
the isospin symmetry. The isoscalar spin-flip M1 strengths were obtained
from the ($d,\,d'$) analysis by assuming that the shape of the
collective transition form factor with the same ${\Delta}J^\pi$ 
is similar in the $^{11}$B($d,\,d'$) and $^{12}$C($d,\,d'$) reactions.
The obtained isovector and isoscalar strengths were used in the DWIA
calculations for the $^{11}$B($p,\,p'$) reaction. The DWIA calculation
reasonably well explains the present $^{11}$B($p,\,p'$) result. However,
the calculated cross section for the 8.92-MeV 3/2$^-_2$ state was
significantly smaller than the experimental values. 
The transition strengths obtained in the shell-model calculations were
found to be 20-50\% larger than the experimental strengths.
The transition strengths for the neutrino induced reactions were
estimated by using the isovector and isoscalar spin-flip M1 strengths.
The present results are quantitatively in agreement with the theoretical
estimation discussing the axial isoscalar coupling in the neutrino
scattering process, and are useful in the measurement of the stellar
neutrinos using the neutral- and charged-current reactions on $^{11}$B.
\end{abstract}

% 21.10.Pc ... Single-particle levels and strength functions
% 25.40.Ep ... Inelastic proton scattering
% 25.45.De ... Elastic and inelastic scattering (2H-induced reactions)
% 25.55.Kr ... Charge-exchange reactions (3H-, 3He-, and 4He-induced reactions)
% 27.20.+n ... 6 <= A <= 19
%\keywords{}

\maketitle

%\section{}
%\subsection{}
%\subsubsection{}

%%%%%%%%%%%%%%%%%%%%%%%%%%%%%%%%%%%%%%%%%%%%%%%%%%%%%%%%%%%%%%%%%%%%%%%
\section{Introduction}

The M1 transition strengths provide important information 
to test the validity of theoretical calculations for nuclear structures.
Recently, the M1 transition strengths are of interest
from the view points of not only nuclear physics but also neutrino
astrophysics because the spin part of the M1 operator is identical with
the relevant operators mediating neutrino-induced reactions. 

\begin{figure}
\includegraphics[scale=0.6]{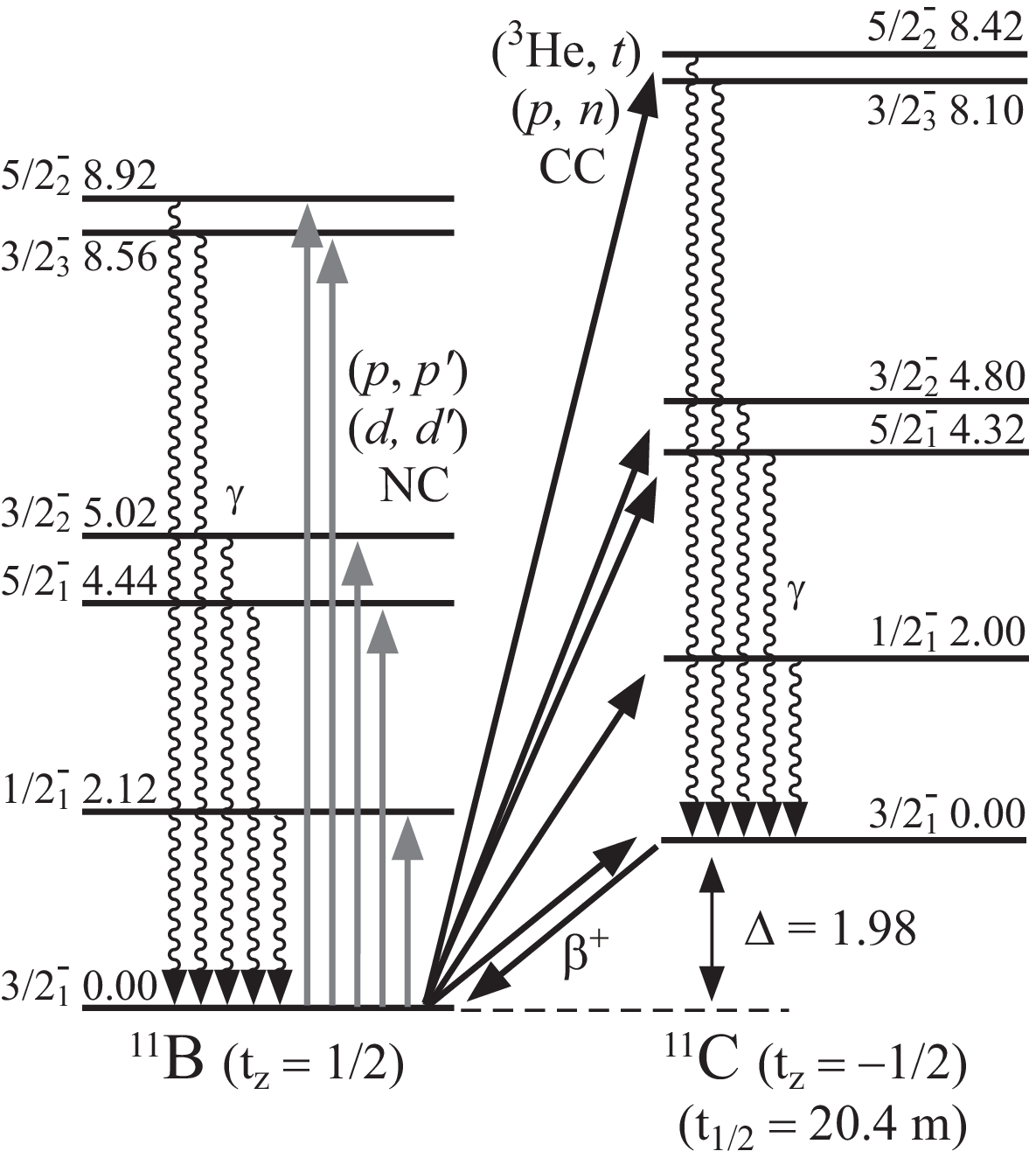}%
\caption{\label{fig:borex}Level scheme for the low-lying states in
 $^{11}$B and $^{11}$C. The states excited by the GT or
 M1 transitions from the ground state of $^{11}$B are shown.}
\end{figure}

Raghavan {\it et al.} pointed out that the $^{11}$B isotope can be used
as a possible neutrino detector to investigate stellar processes
\cite{RAGH86}. High-energy neutrinos emitted from the stellar processes
like the proton-proton fusion chain in the sun and the supernova
explosions
excite low-lying states in $^{11}$B and $^{11}$C by M1 and Gamow-Teller
(GT) transitions via the neutral-current (NC) and charged-current (CC)
processes as seen in Fig.~\ref{fig:borex}. Such neutrinos are detected
by measuring emitted electrons from CC reactions and $\gamma$ rays from
the de-excitations of the low-lying states. Since 
both the NC and CC reactions can be simultaneously measured with one
experimental setup using an isospin
symmetrical relation between the $^{11}$B and $^{11}$C,
the systematic uncertainty in measuring a ratio of the
electron-neutrino flux to the entire neutrino flux is expected to be
small. Since the isospin of the ground state of $^{11}$B is $T=1/2$,
low-lying states in $^{11}$B are excited by both the isovector and
isoscalar transitions. Therefore, both the isovector and
isoscalar responses are needed for estimating the NC cross
section. Bernab\'{e}u {\it et al.} estimated the CC and NC
cross sections for the several low-lying states in $^{11}$B and $^{11}$C
\cite{BERN92}. They obtained the transition strength for the $1/2^-_1$
state from the available experimental data in a model-independent way.
However, the estimations for the other states
rely on the shell-model calculation due to the lack of experimental
data. Thus, the measurement of the M1 strengths for such states in
the $A=11$ system is still important.

In addition to the nuclear response, the coupling constants for the
weak interaction processes are required for estimating the 
cross sections of the neutrino-induced reactions.
Although the axial isovector coupling constant is well-determined as
$g_A=1.254\pm0.006$, the axial isoscalar coupling constant $f_A$ remains
to be
uncertain. The value of $f_A$ directly relates to the strange
quark polarization of nucleon. The EMC experiment reported
$f_A=0.19\pm0.06$ \cite{ASHM89}, but another experiment at SLAC
presented a smaller value of $f_A=0.12\pm0.02$ \cite{ANTH00}. This
discrepancy stems from a difficulty in the deep inelastic scattering
measurement to cover a small-$x$ region. It is noteworthy 
that the strange quark polarization and $f_A$ become attainable by
measuring neutrino-nucleus inelastic scattering with a 
neutrino beam if the nuclear response is precisely determined.

The cross sections for hadronic reactions at forward scattering angles
have a good proportionality to the relevant transition matrix elements.
The cross section for the ($^{3}$He,$\,t$) reaction can be
written in terms of the GT$^-$ operator. On the other hand,
the cross section for the ($d$,$\,d'$) can be described by an isoscalar
M1 operator. The cross section for the ($p$,$\,p'$)
reaction is described by a coherent sum of the 
isovector and isoscalar M1 operators. 
It is, in principle, possible to obtain the GT and M1 strengths 
by comparing the ($^{3}$He,$\,t$), ($d$,$\,d'$), and ($p$,$\,p'$) cross
sections. In this article, we report the results on the GT
and M1 strengths obtained from three experiments on the
$^{11}$B($^3$He,$\,t$),
$^{11}$B($d$,$\,d'$), and $^{11}$B($p$,$\,p'$) reactions.

%%%%%%%%%%%%%%%%%%%%%%%%%%%%%%%%%%%%%%%%%%%%%%%%%%%%%%%%%%%%%%%%%%%%%%%
\section{\label{sec:gtm1}GT and M1 transition operators}

The GT and M1 transition operators are written by
\begin{eqnarray}
\hat{O}({\rm GT}^{\pm})&=&
\frac{1}{2}\sum_{k=1}^{A}{\boldsymbol\sigma}_k\tau_{\pm}(k)\\
\hat{O}(M1)&=&
\left[
\sum_{k=1}^{Z}(g^\pi_l{\bf l}_k+g^\pi_s{\bf s}_k)
+\sum_{k=1}^{N}(g^\nu_l{\bf l}_k+g^\nu_s{\bf s}_k)\right]{\mu_N}
\nonumber\\
&=&
\left[\sum_{k=1}^{A}
\left(
\frac{g^\pi_l+g^\nu_l}{2}{\bf l}_k
+\frac{g^\pi_s+g^\nu_s}{2}\frac{{\boldsymbol\sigma}_k}{2}
\right)
-\left(
\frac{g^\pi_l-g^\nu_l}{2}{\bf l}_k
+\frac{g^\pi_s-g^\nu_s}{2}\frac{\boldsymbol{\sigma}_k}{2}
\right)\tau_z(k)
\right]\mu_N,
\end{eqnarray}
where $\mu_N$ is the nuclear magneton. For protons and neutrons in the
free space, the orbital and spin gyromagnetic factors are 
$g^\pi_l=1$, $g^\nu_l=0$, $g^\pi_s=5.586$, and $g^\nu_s=-3.826$,
respectively. These gyromagnetic factors might be effectively modified
in the nuclear
medium. The eigen values for the isospin operator $\tau_z$ are defined
as $+1$ for neutrons and $-1$ for protons. Using the GT and M1
transition operators, the GT and M1 transition strengths are given by 
\begin{eqnarray}
B({\rm GT}^\pm)&=&\frac{1}{2J_i+1}\left\lvert{\langle}f\lVert
\hat{O}({\rm GT}^\pm){\rVert}i\rangle\right\rvert^2
=\frac{1}{2J_i+1}\left\lvert\frac{1}{2}M(\sigma\tau_\pm)
\right\rvert^2,\\
B(M1)&=&\frac{1}{2J_i+1}\frac{3}{4\pi}
\left\lvert{\langle}f\lVert
\hat{O}(M1){\rVert}i\rangle\right\rvert^2\nonumber\\
&=&\frac{1}{2J_i+1}\frac{3}{4\pi}\left\lvert
g^{IS}_lM(l)+\frac{g^{IS}_s}{2}M(\sigma)
+g^{IV}_lM(l\tau_z)+\frac{g^{IV}_s}{2}M(\sigma\tau_z)
\right\rvert^2\mu_N^2,
\end{eqnarray}
where 
\begin{equation}
g^{IS}_{l(s)}=\frac{g^\pi_{l(s)}+g^\nu_{l(s)}}{2},
\qquad
g^{IV}_{l(s)}=\frac{g^\pi_{l(s)}-g^\nu_{l(s)}}{2},
\qquad
M(\hat{O})={\langle}f\lVert\hat{O}{\rVert}i\rangle.
\end{equation}
$J_i$ is the spin of the initial nuclei. The convention for the reduced
matrix elements is according to that of Edmonds \cite{EDOM60}.
The M1 transition strengths consist of the isovector and isoscalar
parts, and each of them stems from the orbital and spin
contributions. The gyromagnetic factor for the isoscalar spin term is
estimated to be $g^{IS}_s=0.880$ from the free-nucleon
values, which is much smaller than that for the isovector spin term
$g^{IV}_s=-4.706$ in the magnitude. Thus, the isoscalar
spin contribution for the M1 transition strength is 29 times
smaller than the isovector contribution. This means that electro-magnetic
probes are insensitive to the isoscalar spin part. To obtain the
isoscalar spin part, therefore, hadronic probes are useful.

In the case of proton and deuteron scattering off nuclei at
forward angles, the spin part of the M1 transition strength is
dominant. 
The isovector and isoscalar spin-flip M1 strengths are defined by
\begin{eqnarray}
B(\sigma\tau_z)&=&\frac{1}{2J_i+1}\frac{3}{16\pi}
\left\lvert{M(\sigma\tau_z)}\right\rvert^2,\\
B(\sigma)&=&\frac{1}{2J_i+1}\frac{3}{16\pi}
\left\lvert{M(\sigma)}\right\rvert^2.
\end{eqnarray}
There is a simple relation between the isovector
spin-flip M1 strength and the GT strength for the analog transition to
the mirror nucleus under the assumption of the isospin symmetry,
\begin{equation}
\frac{B({\rm GT}^\pm)}{B(\sigma\tau_z)}=\frac{8\pi}{3}
\frac{{\langle}T_i,T_{iz},1,\pm1|T_f,T_{fz}\rangle^2}
{{\langle}T_i,T_{iz},1,0|T_f,T_{fz}\rangle^2}.
\label{eq:gtm1}
\end{equation}
Although the isospin-symmetry breaking changes this ratio, the
deviation is usually small. Therefore, the GT strengths obtained from
the charge exchange reaction are still useful to study the isovector
M1 strengths.

%%%%%%%%%%%%%%%%%%%%%%%%%%%%%%%%%%%%%%%%%%%%%%%%%%%%%%%%%%%%%%%%%%%%%%%
\section{Experiment}

The experiment was performed at the Research Center for Nuclear Physics,
Osaka University, using 450-MeV $^3$He, 200-MeV polarized deuteron, and
392-MeV polarized proton beams. The polarized proton and deuteron beams
were obtained from the high-intensity polarized ion source \cite{HATA97}.
These beams extracted from the ring cyclotron were achromatically
transported to the targets. The beam
intensity on target was in the range of 1-10 enA. A self-supporting
$^{11}$B target with a thickness of 16.7 mg/cm$^2$ and a natural carbon
target with a thickness of 30.0 mg/cm$^2$ were used in the $(d,\,d')$
measurement.
Scattered particles were momentum analyzed by the
high-resolution spectrometer Grand Raiden \cite{MAMO99}. 
The focal-plane detector system of Grand Raiden consisting of
two multiwire drift chambers and plastic scintillation detectors
allowed the reconstruction of the scattering angle at the target via
ray-tracing techniques.
In the forward angle measurements of the ($p$,$\,p'$) and ($d$,$\,d'$)
reactions, a collimator block was placed in front of the focal plane
detectors to avoid a high counting rate due to the elastic
scattering events.  A focal plane polarimeter (FPP) was
used to measure the polarization of protons scattered from the
target in the ($p$,$\,p'$) reaction. The FPP consisted of a
carbon slab with a thickness of 12 cm as a polarization analyzer, four
multiwire proportional chambers, and scintillator hodoscopes
\cite{YOSO95}. For the experimental setup, see
Refs.~\cite{BATA02,AKIM95} and references therein.

\begin{figure}
\includegraphics[scale=0.6]{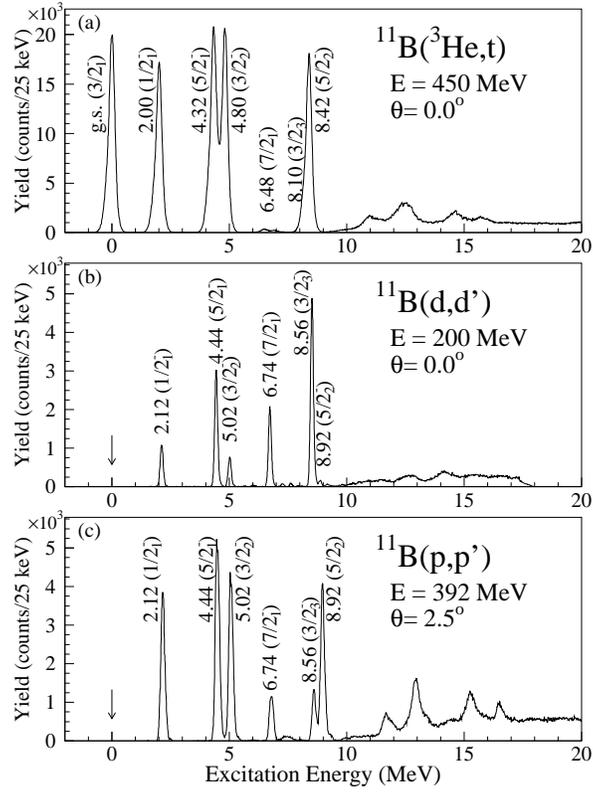}%
\caption{\label{fig:bspe}
Typical spectra for the (a) $^{11}$B($^3$He,$\,t$), (b)
 $^{11}$B($d$,$\,d'$), and (c) $^{11}$B($p$,$\,p'$) reactions.
The peak positions for the elastic deuteron and proton scattering are
indicated by the arrows. See text for details.
}
\end{figure}

Typical spectra of the $^{11}$B($^3$He,$\,t$), $^{11}$B($p$,$\,p'$), and
$^{11}$B($d$,$\,d'$) reactions are shown in Fig.~\ref{fig:bspe}.
In Figs.~\ref{fig:bspe}(b) and (c), elastic scattering events 
disappear since elastically scattered particles were stopped at
the collimator block placed in front of the focal plane detectors.
As seen in Fig.~\ref{fig:bspe}(b), the excitation energy spectrum for
the $^{11}$B($d$,$\,d'$) reaction is observed up to 
$E_x\simeq17.5$ MeV.
An energy resolution of 300 keV full
width at half maximum (FWHM) was obtained in the $^{11}$B($^3$He,$\,t$)
measurement. Since the magnetic spectrometer was operated near the
maximum magnetic field in the $^{11}$B($^3$He,$\,t$) measurement, the
aberration due to the magnetic saturation contributed to deteriorate the
energy resolution. In the $^{11}$B($d$,$\,d'$) and $^{11}$B($p$,$\,p'$)
measurements, energy resolutions of 150 keV and 200 keV (FWHM) were
obtained respectively, which were dominated by the energy spreads of the
cyclotron beams. All the prominent peaks were identified as those of
known states in $^{11}$B or $^{11}$C \cite{AJZE90}. Contaminating
impurities in the $^{11}$B target were identified by the kinematic
energy shift in the elastic scattering at backward angles, and those
contributions were estimated to be less than 1\%, 0.8\%, and 0.08\% for
$^{10}$B, $^{12}$C, and $^{16}$O, respectively.

%%%%%%%%%%%%%%%%%%%%%%%%%%%%%%%%%%%%%%%%%%%%%%%%%%%%%%%%%%%%%%%%%%%%%%%
\section{Result and discussion}

Since the spin-isospin term $V_{\sigma\tau}$ in the effective
interaction is dominant in the
($^3$He,$\,t$) reaction at 150 MeV/nucleon, transitions to the strong
peaks in Fig.~\ref{fig:bspe}(a) are inferred to have a spin-flip
nature. On the other hand, the scalar term $V_0$ is dominant in the
($d$,$\,d'$) reaction at 100 MeV/nucleon. Thus, it is natural to infer
that the strongly excited states in
Fig.~\ref{fig:bspe}(b) have a non-spin-flip nature. This
simple consideration leads to a qualitative conclusion that
the 7/2$^-_1$ and 3/2$^-_3$ states
have a non-spin-flip characteristic while the 5/2$^-_2$ state mainly
has a spin-flip characteristic. 

%%%%%%%%%%%%%%%%%%%%%%%%%%%%%%%%%%%%%%%%%%%%
\subsection{\label{sub:het}$^{11}$B($^3$He$,\,t$) reaction}

\begin{figure}
\includegraphics[scale=0.7]{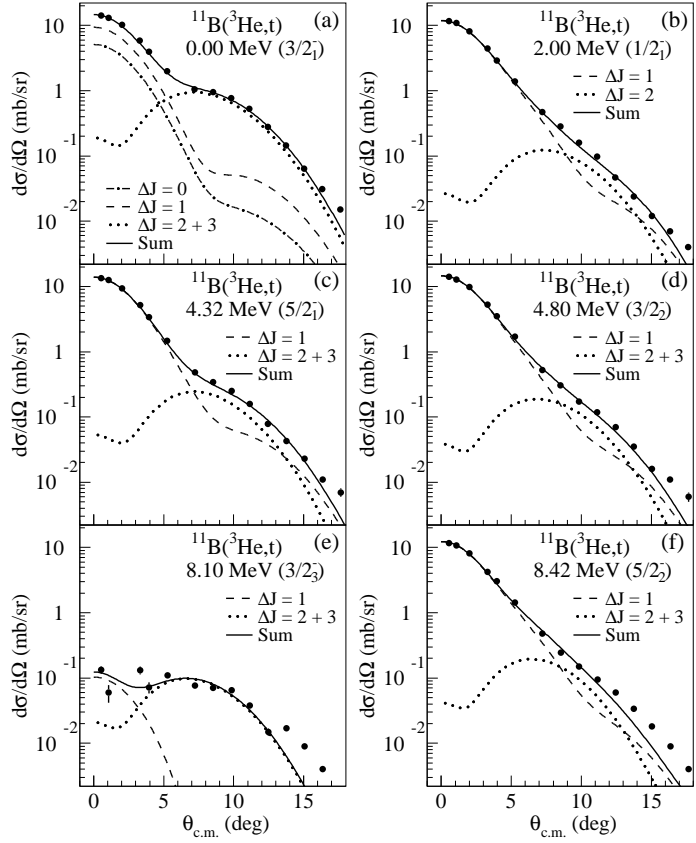}%
\caption{\label{fig:het}
Cross sections for the  $^{11}$B($^3$He,$\,t$) reactions compared with
the DWIA calculation.
The dash-dotted, dashed, and dotted curves show the ${\Delta}J=0$,
${\Delta}J=1$, and ${\Delta}J\ge2$ contributions, respectively. The
solid curves are the sums of all the multipole contributions. }
\end{figure}

The cross sections for the $^{11}$B($^3$He,$\,t$) reaction are shown
in Fig.~\ref{fig:het}. To determine the GT strength from the
cross sections, the distorted-wave impulse-approximation (DWIA)
calculation was performed using a computer code DWBA98
\cite{RAYN99}. 
Optical-model parameters for the
entrance channel were obtained from the $^3$He elastic scattering on
$^{13}$C at $E(^3{\rm He})=450$ MeV \cite{CHIBI04}. For the exit channel,
the same radius and diffuseness with the entrance channel were used
while the potential depth was modified to be 85\% of the depth for the
entrance channel \cite{WERF89}. Wave functions were obtained from the
shell-model calculation using the POT interaction by Cohen and
Kurath \cite{COHE65}. Single-particle wave functions were calculated
by using a harmonic oscillator (HO) potential, and the oscillation
length $b$ for the HO potential was determined for each state to
reproduce the measured cross sections as tabulated in
Table~\ref{tab:bgt}. 

An effective $^3$He-N interaction with isospin ($V_\tau$), spin-isospin
($V_{\sigma\tau}$), and
isospin-tensor ($V^T_\tau$) terms, represented by a Yukawa potential,
was employed to describe the projectile-target interaction. 
Ranges for the Yukawa potentials were fixed at 1.42 fm and 0.88 fm for
the central and tensor terms, respectively. For the central potential,
the strength ratio of $R^2=|V_{\sigma\tau}/V_\tau|^2=8.24\pm0.11$ is
widely used at 450 MeV \cite{AKIM95}, but recent studies implies
that a smaller value of $R^2$ might be preferable for light nuclei
\cite{FUJI03}. Actually, the $R^2$ value is estimated to be 5.24 at
150~MeV/nucleon and A=11 from the Franey-Love interaction \cite{FRAN85}.
We, therefore, tested both
$R^2=8.24$ and $R^2=5.24$ to study the effect on the final $B({\rm GT}^-)$
values by $R^2$. 

The absolute strength of the central potential was
determined by comparing the DWIA calculation with the experimental data
for the ground-state transition. The cross section was described by an
incoherent sum over the cross sections for the different multipole
transitions,
\begin{equation}
\frac{d\sigma}{d\Omega}=\sum_{{\Delta}J^\pi}A({\Delta}J^\pi)
\frac{d\sigma}{d\Omega}({\Delta}J^\pi),
\end{equation}
where the normalization factors $A({\Delta}J^\pi)$'s were determined to
reproduce the experimental data. Since each multipole cross section
was calculated by using shell-model wave functions, the GT and Fermi
transition strengths were related to the normalization factors by
\begin{eqnarray}
B({\rm GT}^-)_{\rm exp}&=&A({\Delta}J^\pi=1^+)B({\rm GT}^-)_{\rm SM},\\
B({\rm F})_{\rm exp}&=&A({\Delta}J^\pi=0^+)B({\rm F})_{\rm SM},
\end{eqnarray}
where $B({\rm GT}^-)_{\rm SM}$ and $B({\rm F})_{\rm SM}$ are the GT and
Fermi strengths predicted by the shell-model calculation.
The ground-state GT transition strength is known
to be $B({\rm GT}^-)=0.345\pm0.008$ from the $\beta$-decay strength, and
the Fermi transition strength is given by $B({\rm F})=N-Z=1$. Since the
POT wave function predicted $B({\rm GT}^-)=0.623$ and
$B({\rm F})=1.000$ for
the ground-state transition, the normalization factors were fixed at
$A({\Delta}J^\pi=1^+)=0.554$ and $A({\Delta}J^\pi=0^+)=1.000$ in the
analysis. Finally, $V_{\tau}=1.5$ MeV and $V_{\sigma\tau}=-4.2$ MeV were
obtained for $R^2=8.24$ while $V_{\tau}=1.7$ MeV and
$V_{\sigma\tau}=-3.9$ MeV were for $R^2=5.24$. The value of $V^T_\tau$
was determined to be $-2.5$ MeV/fm$^2$. 

The calculated cross sections for the transitions to the ground and
excited states in $^{11}$C are compared with the experimental results in
Fig.~\ref{fig:het}. 
Although only the results of
the calculations with $R^2=8.24$ are shown in Fig.~\ref{fig:het}, those
of the calculations with
$R^2=5.24$ are quite similar. As seen in Fig.~\ref{fig:het}, the DWIA
calculations successfully explain the experimental results. 

The GT strengths
obtained from the ${\Delta}J^\pi=1^+$ cross sections are compared with
the previous $(p,\,n)$ result \cite{TADD90} in Table~\ref{tab:bgt}. 
The present results are consistent with the $(p,\,n)$ result although
several states are not resolved in the $(p,\,n)$ measurement due to poor
energy resolution. We found that the $R^2=8.24$
results give the $B({\rm GT}^-)$ values close to the $(p,\,n)$ result.
The $B({\rm GT}^-)$ values for
$R^2=8.24$ are smaller than those for
$R^2=5.24$ by 15\%. This difference of 15\%
in the $B({\rm GT}^-)$ values almost equals to the difference in
the $|V_{\sigma\tau}|^2$ values between the $R^2=8.24$ and 5.24
cases. This fact indicates that the factorized expression for the charge
exchange reaction \cite{TADD87} gives a good description for the
($^3$He,$\,t$) reaction
at $E(^3{\rm He})=450$ MeV. The $B({\rm GT}^-)$ value for the 8.10-MeV
$3/2^-_3$ state is very small and this result is consistent with the
previous simple consideration that the $3/2^-_3$ state has a
non-spin-flip characteristic.

\begin{table}
\caption{\label{tab:bgt}Oscillation length used in the DWIA calculation
and $B({\rm GT})$ values.}
\begin{ruledtabular}
\begin{tabular}{cccccc}
$E_x$ (MeV) & $J^\pi$ & $b$ (fm) &
 \multicolumn{2}{c}{$B({\rm GT}^-)$\footnote{Present result.}} &
$B({\rm GT}^-)$\footnote{From Ref.~\cite{TADD90}} \\
 & & & $R^2=8.24$ & $R^2=5.24$ & \\
\hline
0.00 & $3/2^-$ & 1.61 & \multicolumn{2}{c}{$0.345\pm0.008$} & $0.345\pm0.008$\\
2.00 & $1/2^-$ & 1.61 & $0.402\pm0.031$ & $0.461\pm0.036$ & $0.399\pm0.032$ \\
\begin{tabular}{c} 4.32            \\ 4.80            \\ \end{tabular}&
\begin{tabular}{c} $5/2^-$         \\ $3/2^-$         \\ \end{tabular}&
\begin{tabular}{c} 1.66            \\ 1.73            \\ \end{tabular}&
\begin{tabular}{c} $0.454\pm0.026$ \\ $0.480\pm0.031$ \\ \end{tabular}&
\begin{tabular}{c} $0.521\pm0.031$ \\ $0.551\pm0.036$ \\ \end{tabular}&
$\Bigr\}$\hspace{4pt} $0.961\pm0.060$ \hspace{6pt}\mbox{ } \\
\begin{tabular}{c} 8.10            \\ 8.42            \\ \end{tabular}&
\begin{tabular}{c} $3/2^-$         \\ $5/2^-$         \\ \end{tabular}&
\begin{tabular}{c} 1.81            \\ 1.88            \\ \end{tabular}&
\begin{tabular}{c} $\le 0.003$     \\ $0.406\pm0.038$ \\ \end{tabular}&
\begin{tabular}{c} $\le 0.004$     \\ $0.466\pm0.045$ \\ \end{tabular}&
$\Bigr\}$\hspace{4pt} $0.444\pm0.010$ \hspace{6pt}\mbox{ }\\
\end{tabular}
\end{ruledtabular}
\end{table}

%%%%%%%%%%%%%%%%%%%%%%%%%%%%%%%%%%%%%%%%%%%%
\subsection{$^{11}$B($d$,$\,d'$) and $^{12}$C($d$,$\,d'$) reactions}

\begin{figure}
\includegraphics[scale=0.5]{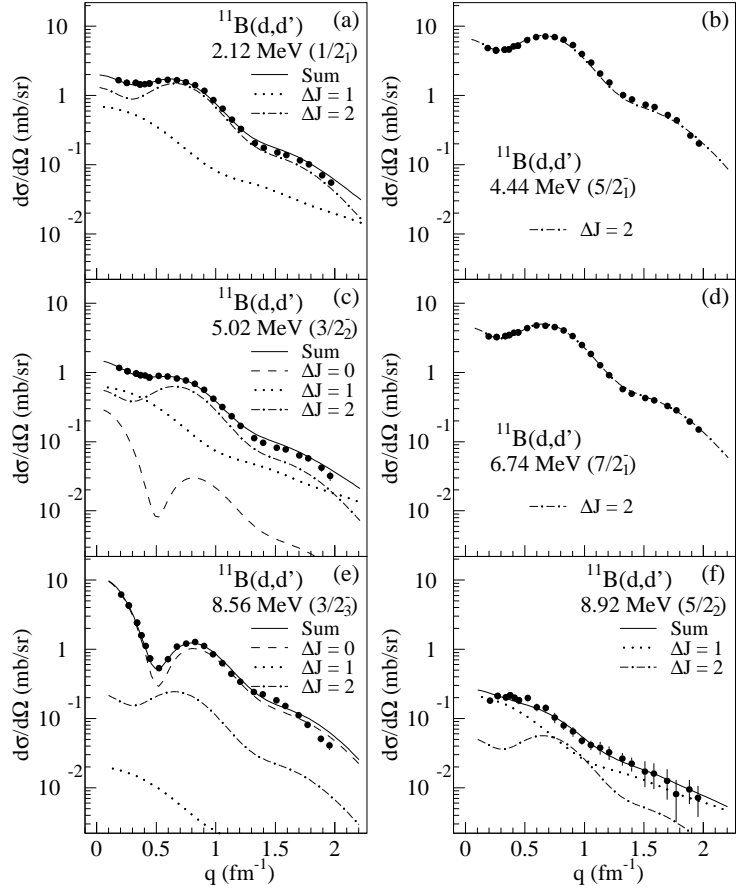}%
\caption{\label{fig:b11dd}
Cross sections for the non-parity changing transitions in deuteron
inelastic scattering on $^{11}$B.
The dashed, dotted, and dash-dotted curves show the ${\Delta}J=0$,
1, and 2 contributions, respectively. The
solid curves are the sums of all the multipole contributions. }
\end{figure}

\begin{figure}
\includegraphics[scale=0.7]{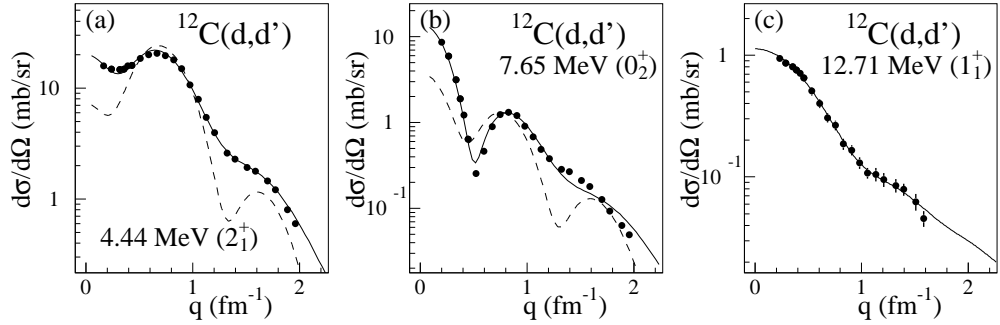}%
\caption{\label{fig:c12dd}
Cross sections for the non-parity changing transitions in deuteron
inelastic scattering on $^{12}$C.
}
\end{figure}

The measured cross sections for the $^{11}$B($d$,$\,d'$) and
$^{12}$C($d$,$\,d'$)
reactions exciting the several low-lying states are shown in
Figs.~\ref{fig:b11dd} and \ref{fig:c12dd}.
The $(d,\,d')$ cross section was
given by an incoherent sum over the cross sections for the different
multipole transitions. To derive the isoscalar spin-flip M1 strength
$B(\sigma)$ from the experiment, the $(d,\,d')$ cross section for each
${\Delta}J^\pi$ transition was needed.

Microscopic calculations for deuteron induced reactions are generally
difficult because the deuteron is a loosely bound two-body system and 
its internal degree of freedom has to be taken into account. Recently, a
microscopic calculation using a three-body $d$-$N$ interaction was
performed and gave a good description for the $^{12}$C($d$,$\,d'$)
reaction at $E_d=270$ MeV at backward angles of $\theta\ge 5^\circ$
\cite{SATO02}. This calculation, however, does not reproduce the angular
distribution of the cross section at forward angles near $0^\circ$ where
${\Delta}J^\pi=1^+$ transitions become strong. Therefore, the
application of such microscopic calculations is not suitable for our
purpose to estimate the cross sections for each ${\Delta}J^\pi$
transition and to determine the
isoscalar spin-flip M1 strength. In the present work, we measured the
$^{12}$C($d$,$\,d'$) reaction for the comparison with the
$^{11}$B($d,\,d'$) reaction. Since the spin-parity of the ground state
of $^{12}$C is $0^+$, transitions to the discrete states in $^{12}$C
are expected
to be good references for the angular distributions of the cross sections
for certain ${\Delta}J^\pi$ transitions.

\begin{figure}
\includegraphics[scale=0.6]{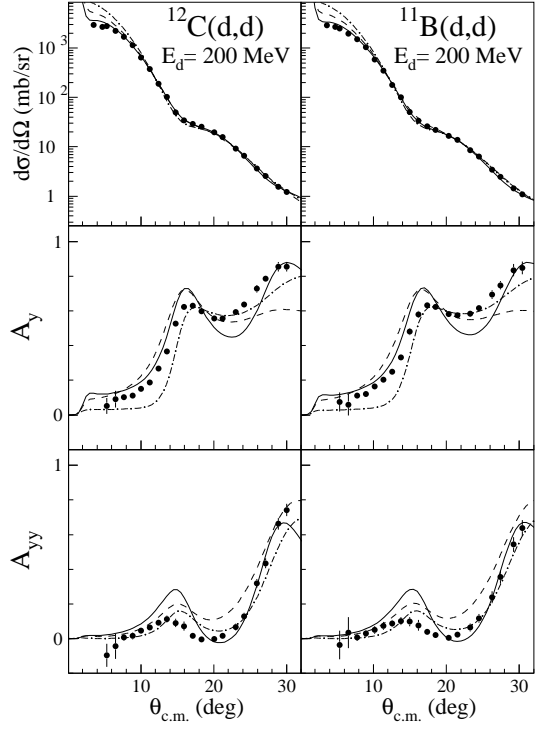}%
\caption{\label{fig:ddela}
Cross sections, vector analyzing powers $A_y$, and tensor analyzing powers
 $A_{yy}$ for the deuteron elastic scattering on $^{12}$C and
 $^{11}$B compared with the results of the optical-model calculations.
The sold lines show the results with the parameter set
'BC'. The dashed and dash-dotted lines show the results using the
 modified optical-model potentials for the
${\Delta}J^\pi=0^+$ and $2^+$ transitions, respectively.
}
\end{figure}

To parameterize the angular distributions of the cross sections for 
the 4.44-MeV $2^+_1$ and 7.65-MeV $0^+_2$ states in
$^{12}$C, we asked for a help of a deformed potential (DP) model
calculation using a computer code ECIS95 \cite{RAYN95}. Optical-model
potentials for the DP-model analysis were parameterized as
\begin{equation}
U(r)=-Vf(r,r_R,a_R)-iWf(r,r_I,a_I)
+2\left(\frac{\hbar}{{m_\pi}c}\right)^2V_{SO}\frac{1}{r}\frac{d}{dr}
f(r,r_{SO},a_{SO}){\boldsymbol l}\cdot{\boldsymbol s}+V_C,
\label{eq:opt}
\end{equation}
where
\begin{equation}
f(r,r_x,a_x)=\left[1+\exp\left(\frac{r-r_xA^{1/3}}{a_x}\right)\right]^{-1}.
\end{equation}
The Coulomb potential $V_C$ was taken as that of a uniformly charged
sphere with a radius of $R_C=1.3A^{1/3}$~fm. Three parameter sets 'B', 'C',
and 'BC' tabulated in Table~\ref{tab:opt} were determined by fitting the
cross sections, vector analyzing powers ($A_y$), and tensor analyzing
powers ($A_{yy}$) for the elastic scattering from the $^{11}$B and
$^{12}$C targets. The parameter sets 'B' and 'C'
were obtained from the independent analysis of the $^{11}$B and $^{12}$C
data. The parameter set 'BC' was determined by fitting both the $^{11}$B
and $^{12}$C data simultaneously. As seen from the $\chi^2$ values in
Table~\ref{tab:opt}, almost the same quality of the fit is
obtained for the three parameter sets. The results from the
optical-model calculation with the parameter set 'BC' are presented by
the solid lines in Fig.~\ref{fig:ddela}.

Using the obtained
optical-model potential, we calculated the transition potential according
to the prescription of the DP model \cite{SATC83}. The result of the 
DP-model calculation with the parameter set 'BC' is shown by the
dashed lines in Figs.~\ref{fig:c12dd}(a) and (b). Unfortunately the 
DP-model calculation with the parameter set 'BC' is not satisfactory for
reproducing
the cross sections for the $2^+_1$ and $0^+_2$ states. To obtain the
better descriptions, we modified the optical-model
potentials by fitting the elastic and inelastic scattering data
simultaneously. During the process of the modification, the potential
parameters for the ${\Delta}J^\pi=0^+$ and $2^+$ transitions were
searched for independently, and we additionally introduced the real and
imaginary surface terms into the optical-model potential. This procedure
means that we allocate additional degrees of freedom for the
precise parameterization of the potential shape in the surface region
where the incident deuterons are strongly absorbed. As the result of the
modification, the ${\Delta}J^\pi=2^+$ and
$0^+$ cross sections were well-fitted as shown by the solid lines in
Figs.~\ref{fig:c12dd}(a) and (b) although the optical-model calculations
for the elastic scattering became slightly worse as seen in
Fig.~\ref{fig:ddela}. 
We estimated that the change of the cross-section values due to the
coupled channel effect is smaller than 10\%. This effect is, therefore,
neglected in the present analysis. 

Although a question is addressed on the physical
interpretation of the potential
modification described above, it should be emphasized that our purpose
is simply to parameterize the angular distributions of the cross section
for the ${\Delta}J^\pi=0^+$ and $2^+$ transitions in the $^{12}$C($d,\,d'$)
reaction and to perform the multipole decomposition analysis for the
$^{11}$B($d,\,d'$) reaction. The basic assumption of the present analysis
is that the collective transition form factors reflected to the
cross-section shape for a certain ${\Delta}J^\pi$ transition in the
$^{12}$C($d,\,d'$) and $^{11}$B($d,\,d'$) reactions are similar. The
validity of this assumption could be tested by checking whether the
angular distribution in the $^{11}$B($d,\,d'$) reaction is reasonably
explained or not.

\begin{table}
\caption{\label{tab:opt}Optical model parameters derived from our best
fit to the elastic scattering data. Parameter sets 'B' and 'C' were
determined by fitting the elastic scattering on $^{11}$B and $^{12}$C
respectively, while 'BC' was by fitting both the $^{11}$B and $^{12}$C
data simultaneously.
}
\begin{ruledtabular}
\begin{tabular}{ccccccccccc}
    & $V$ & $r_R$ & $a_R$ & $W$ & $r_I$ & $a_I$ & $V_{SO}$ 
    & $r_{SO}$ & $a_{SO}$ & $\chi^2$\\
    & (MeV) & (fm)  & (fm)  & (MeV) & (fm)  & (fm)  & (MeV)
    & (fm)  & (fm)  &  \\
\hline
{\rm B} & 20.36 & 1.50 & 0.65
        & 35.25 & 0.66 & 1.06
        & 1.68  & 1.00 & 0.68 & 1160 \\
{\rm C} & 20.45 & 1.51 & 0.65
        & 31.45 & 0.78 & 1.02
        & 1.79  & 0.98 & 0.70 & 1396 \\
{\rm BC}& 20.24 & 1.51 & 0.65
        & 33.27 & 0.72 & 1.05
        & 1.74  & 0.99 & 0.69 & 2599 \\
\end{tabular}
\end{ruledtabular}
\end{table}

On the other hand, the angular distribution of the cross section for the
12.71-MeV $1^+_1$ state was parameterized as a
function of the momentum transfer $q$ using spherical Bessel functions
\begin{equation}
\frac{d\sigma}{d\Omega}(1^+_1)=
\alpha_0\left|j_0(qR_0)\right|^2+\alpha_2\left|j_2(qR_2)\right|^2.
\end{equation}
Four parameters in the above equation were determined to be 
$R_0=1.26A^{1/3}$~fm, $R_1=1.52A^{1/3}$~fm, $\alpha_0=1.13$~mb/sr, and
$\alpha_1=1.27$~mb/sr by fitting the experimental data as shown by the
solid line in Fig.~\ref{fig:c12dd}(c).

The cross sections for the ${\Delta}J^\pi=0^+$, $1^+$ and $2^+$
transitions in the $^{11}$B($d,\,d'$) reaction were calculated using the
modified potentials for the DP-model calculation and the spherical
Bessel functions which were determined in the $^{12}$C($d,\,d'$)
analysis. Then, the measured $^{11}$B($d,\,d'$) cross sections were
fitted by combining the calculated cross sections. In the fitting
procedure, the
higher multipole contributions with ${\Delta}J\ge3$ were neglected. As
seen in Fig.~\ref{fig:b11dd}, the experimental results were well
reproduced in the fit. This means that the basic assumption on the
similarity between the $^{12}$C($d,\,d'$) and $^{11}$B($d,\,d'$)
reactions is reasonable.

Although the 4.44-MeV $5/2^-_1$ state can be excited by both the
${\Delta}J^\pi=1^+$ and $2^+$ transitions, the main part of the
transition is due to ${\Delta}J^\pi=2^+$.
This result is explained if 
the ground and 4.44-MeV states are the strongly coupled members 
of the ground-state rotational band. 
Since the observed ${\Delta}J^\pi=2^+$ transition strength is
much stronger than the expected ${\Delta}J^\pi=1^+$ strength, the
${\Delta}J^\pi=1^+$ component of the transition strength can not be
reliably extracted for the 4.44-MeV state. The transition strength for
the 6.74-MeV $7/2^-_1$ state is also dominated by the
${\Delta}J^\pi=2^+$ component although the ${\Delta}J^\pi=1$ transition
to this state is not allowed.

The cross section for the ${\Delta}J^\pi=1^+$ transition is known to be
proportional to the isoscalar
spin-flip M1 strength $B(\sigma)$. The $B(\sigma)$ value for the
transition to the 2.12-MeV $1/2^-_1$ state is 
$0.037\pm0.008$, which is obtained from the $\gamma$-decay widths of the
mirror states and
the $B({\rm GT}^-)$ value \cite{BERN92}. Using this value, the cross
sections for the ${\Delta}J^\pi=1^+$ transitions for other excited
states were
converted to the $B(\sigma)$ values as tabulated in Table~\ref{tab:bsigma}.
Systematic uncertainties on $B(\sigma)$ are mainly due to errors in the
model calculation for the ($d,\,d'$) reaction. The normalization
uncertainty of 20\% is not shown in Table~\ref{tab:bsigma}, which is
attributed to the uncertainty on the $B(\sigma)$ value for the 2.12-MeV
state taken from Ref.~\cite{BERN92}.

\begin{table}
\caption{\label{tab:bsigma}
Isoscalar spin-flip M1 transition strength $B(\sigma)$. 
Systematic uncertainties are mainly due to errors in the DP model
calculations. The normalization uncertainty of 20\% attributed to the
uncertainty on the calibration reference is not shown.}
\begin{ruledtabular}
\begin{tabular}{ccc}
$E_x$ (MeV) & $J^\pi$ & $B(\sigma)$ \\
\hline
2.12 & $1/2^-_1$ & $0.037\footnote{From Ref.~\cite{BERN92}}\pm0.007$ \\
4.44 & $5/2^-_1$ & $-$ \\
5.02 & $3/2^-_2$ & $0.035\pm0.005$ \\
8.56 & $3/2^-_3$ & $\le0.003$ \\
8.92 & $5/2^-_2$ & $0.012\pm0.003$ \\
\end{tabular}
\end{ruledtabular}
\end{table}

The 8.56-MeV $3/2^-_3$ state, which is not predicted by the 
shell-model calculation with the POT potential, has a strong
${\Delta}J^\pi=0^+$ component. Although the isobaric analog state of the
8.56-MeV state is observed at $E_x=8.10$ MeV in the
$^{11}$B($^3$He$,\,t$) reaction, the excitation strength is extremely
weak. These facts indicate that the $3/2^-_3$ state in the $A=11$ system
is collective, and the single-particle aspect of its wave function is
small.

It is noteworthy that the $A=11$ and 12 systems have several
similarities in the collective nature, {\it i.e.} the first member of
the ground-state rotational band appears at $E_x=4.44$ MeV and the
collective E0 state appears around $E_x=8$ MeV. Although the deeper
discussion about the collective nature of $^{11}$B is out of the scope
of the present work, a comparative study of the collective nature in the
$A=11$ and 12 systems might provide precious information to improve the
nuclear cluster model which is one of recent interesting topics
\cite{HORI04}.

%%%%%%%%%%%%%%%%%%%%%%%%%%%%%%%%%%%%%%%%%%%%
\subsection{\label{sec:pp} $^{11}$B($p,\,p'$) reaction}

\begin{figure}
\includegraphics[scale=0.8]{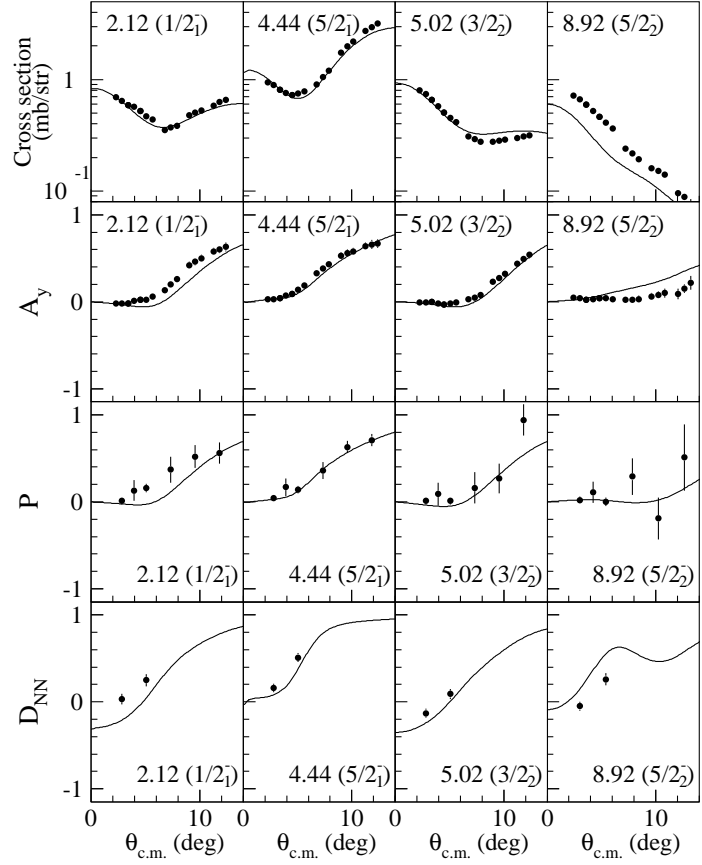}%
\caption{\label{fig:pp}
Cross sections, vector analyzing powers $A_y$, induced polarizations $P$,
and depolarization parameters $D_{NN}$ for the $^{11}$B($p$,$\,p'$)
reaction compared with the results of the DWIA calculations.
}
\end{figure}

The cross sections, analyzing powers ($A_y$), induced polarizations ($P$),
and depolarization parameters ($D_{NN}$) for the $^{11}$B($p$,$\,p'$)
reaction are shown in Fig.~\ref{fig:pp}. 
The DWIA calculation was performed by using the
computer code DWBA98. The
effective nucleon-nucleon interaction derived by Franey and Love at 425
MeV \cite{FRAN85} was used in the calculations. The global Dirac
optical-model potential was used in the Shr\"{o}dinger equivalent form
\cite{HAMA90} to give the distorted waves of incoming and outgoing
protons. The ($p,\,p'$) cross sections
were given by an incoherent sum over the cross section of the different
multipole contribution as in the cases of the analyses for the
($^3$He$,\,t$) and ($d,\,d'$) reactions. The cross section for each
${\Delta}J^\pi$ transition was
described by a coherent sum of the isovector and isoscalar
contributions. 

Since the isovector and isoscalar transition strengths
for the several states in $^{11}$B have been already obtained from the
($^3$He$,\,t$) and ($d,\,d'$) analyses in the present work, the DWIA
calculation for the $^{11}$B($p,\,p'$) reaction can be
performed without any free parameters. The isovector strengths
determined by using $R^2=8.24$ were used in the calculation. 
Since the isoscalar spin-flip M1 strength was
not reliably determined in the ($d,\,d'$) analysis for the 4.44-MeV
state, the isoscalar spin-flip M1 strength was assumed to be
\begin{equation}
B(\sigma)_{\rm exp}=B(\sigma\tau_z)_{\rm exp}\,
\frac{B(\sigma)_{\rm SM}}{B(\sigma\tau_z)_{\rm SM}},
\label{eq:bsigma}
\end{equation}
{\it i. e.} the relative strength of the isoscalar
transition to the isovector transition was taken from the the 
shell-model calculation.

As shown in Fig.~\ref{fig:pp}, the
DWIA calculation well explains the experimental results except the
cross section for the transition to the 8.92-MeV state. This supports the
reliability of the $^{11}$B($^3$He$,\,t$) and $^{11}$B($d,\,d'$) analyses.
The calculated cross section for the 8.92-MeV state is about 50\%
smaller than the experiment. One remarkable fact predicted by the DWIA
calculation using the POT wave functions is that the isovector
and isoscalar components destructively contribute for the
${\Delta}J^\pi=1^+$ transition to the 8.92-MeV state, while
these two components constructively contribute for the 2.12-MeV,
4.44-MeV, and 5.02-MeV states. However, the considerable decrease of
50\% in the ($p,\,p'$) cross section can not be explained even if
a constructive interference is assumed.
The reason why the DWIA
calculation underestimates the cross section for the transition to the
8.92-MeV state is still unclear. The result may imply that the 8.92-MeV
state has a wave function totally different from the other states
at 2.12, 4.44, and 5.02 MeV.

Since
the $\left|V_{\sigma\tau}/V_{\sigma}\right|^2$ value of the effective
nucleon-nucleon interaction is large, the cross section for the
${\Delta}J^\pi=1^+$ transition is dominated by the isovector component
and is insensitive to the relative strength between the isovector and
isoscalar components. However, the polarization transfer observables are
sensitive to the relative strength in the spin-flip transition. Since
the depolarization parameter was measured in the present experiment, the
relative strength between the isovector and isoscalar components in the
${\Delta}J^\pi=1^+$ transition is, in principle, determined from the
$(p,\,p')$ result. Although we tried to determine the relative
strength by fitting the measured depolarization parameter, such analyses
were found to be unacceptable; the isoscalar strengths for all the
states are strongly suppressed. This unrealistic result is mainly caused
by errors in the effective nucleon-nucleon interaction. Therefore, the
precise determination of the effective interaction is strongly
desired. Additional measurements of other kinds of the polarization
transfer observables ({\it e.g.} $D_{SS}$ and $D_{LL}$) would be helpful
in clarifying the unclear situation discussed above.

%%%%%%%%%%%%%%%%%%%%%%%%%%%%%%%%%%%%%%%%%%%%
\subsection{Comparisons with the previous experiment and
the shell-model calculation}

\begin{figure}
\includegraphics[scale=0.5]{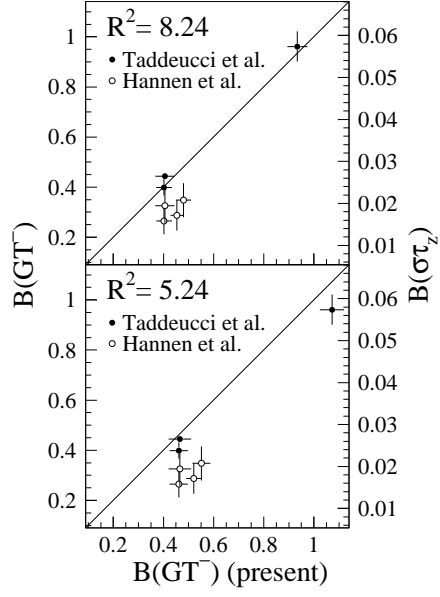}%
\caption{\label{fig:compexp}
$B({\rm GT}^-)$ from Ref.~\cite{TADD90} (solid circles) and
$B(\sigma\tau_z)$ from Ref.~\cite{HANN03} (open circles) are compared
with the present results in the cases of $R^2=8.24$ (upper panel) and
$R^2=5.24$ (lower panel). 
The horizontal axis indicates the $B({\rm GT}^-)$ values from the present
study and the vertical axis shows the $B({\rm GT}^-)$ or
$B(\sigma\tau_z)$ values from the previous experiments.
The solid lines are drawn to guide the eye.
}
\end{figure}

Although the two doublets at 4.32-4.80 MeV and 8.11-8.42 MeV in $^{11}$C
are not separately resolved in the previous ($p,\,n$) measurement
\cite{TADD90}, the $B({\rm GT}^-)$ values for the charge exchange
reactions to the low-lying
states in $^{11}$C have been obtained.
Recently, the $B(\sigma\tau_z)$ values for the low-lying
states in $^{11}$B are reported by KVI group from the ($p,\,p'$)
measurement at $E_p=150$~MeV \cite{HANN03}. These results are compared
with the present results in
Fig.~\ref{fig:compexp}. 
The result with $R^2=8.24$ is preferable from a view of
the consistency with the ($p,\,n$) results. The results from
Ref.~\cite{HANN03} are systematically smaller than those from the other
experiments. This inconsistency seems to be originated from the
difficult normalization process in the DWIA calculation. 
The calculated cross sections depend on various parameters,
{\it i.e.} the nuclear transition strength, the distorting potential,
the effective interaction, and so on. The uncertainties of the
distorting potential and the effective interaction cause large
systematic uncertainties in the normalization. 

\begin{figure}
\includegraphics[scale=0.5]{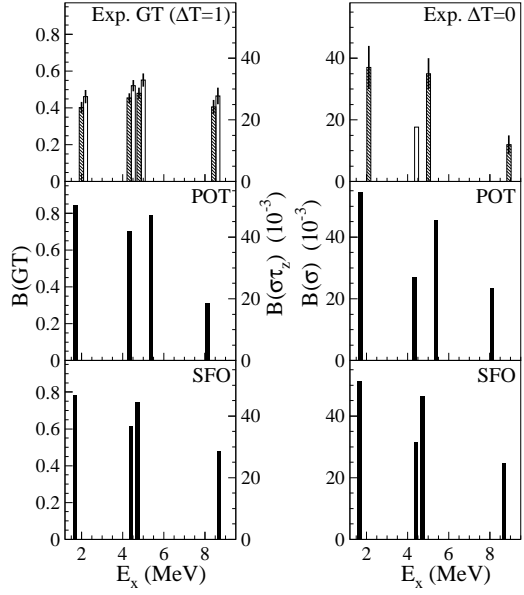}%
\caption{\label{fig:smexp}
Measured $B({\rm GT}^-)$ ($B(\sigma\tau_z)$) and $B(\sigma)$ values are
compared with the shell-model predictions using the POT
\cite{COHE65} and SFO \cite{SUZU03} interactions. The hatched and open
bars in the left panels show the $B({\rm GT})$ results deduced by using
$R^2=8.24$ and $R^2=5.24$, respectively. The open bar in the right panel
shows the $B(\sigma)$ value for the 4.44-MeV state estimated from 
$B({\rm GT})$ (see text).
}
\end{figure}

The present results are compared with the shell-model predictions using
the POT \cite{COHE65} and SFO (Suzuki-Fujimoto-Otsuka)
\cite{SUZU03} interactions in Fig.~\ref{fig:smexp}. The shell-model
calculations with the POT and SFO interactions are performed in
the $0\hbar\omega$ and $0-2\hbar\omega$ configuration spaces,
respectively. The hatched and open bars in the left panels show the
$B({\rm GT}^-)$ results deduced from the analyses using $R^2=8.24$ and
$R^2=5.24$, respectively. The open bar in the upper-right panel shows
the $B(\sigma)$ value for the 4.44-MeV state, which is estimated
from the $B({\rm GT}^-)$ value by using Eqs.~(\ref{eq:gtm1}) and
(\ref{eq:bsigma}). Both the calculations with
the POT and SFO wave functions well explain the
excitation energies, but the predicted excitation strengths are
unacceptably large. The calculations with the SFO wave functions are
better in describing the excitation strength since the 2p-2h
configuration mixing is taken into account, but the excitation strengths
are still overestimated by 20-50\%.

%%%%%%%%%%%%%%%%%%%%%%%%%%%%%%%%%%%%%%%%%%%%
\subsection{Transition strengths for neutrino induced reactions}

The cross section for the CC transition $^{11}$B($\nu_e,\,e^-$) is given
as follows at the long wavelength limit \cite{BERN92},
\begin{equation}
\sigma_{CC}=\frac{(G_F\cos\theta_c)^2}{4\pi}\frac{1}{2J_i+1}
\left[\left|M(\tau_-)\right|^2+g_A^2\left|M(\sigma\tau_-)\right|^2\right]
E'_e\left|{\bf p}'_e\right|F(Z, E'_e),
\end{equation}
where $G_F$ is the Fermi coupling constant, $\theta_c$ is the Cabibbo
angle, $g_A$ is the axial isovector coupling constant, $J_i$ is the
initial nuclear spin, $E'_e$ (${\bf p}'_e$) is the energy (momentum) of
the outgoing electron, $F(Z,E'_e)$ is the Coulomb function for the final
electron, and $Z$ is the charge of the residual nucleus. The
cross section for the NC transition $^{11}$B($\nu_x,\,\nu'_x$) is
\begin{equation}
\sigma_{NC}=\frac{(G_FE')^2}{4\pi}\frac{1}{2J_i+1}
\left|f_AM(\sigma)+g_AM(\sigma\tau_z)\right|^2,
\end{equation}
where $f_A$ and $E'$ are the axial isoscalar coupling constant and
outgoing neutrino energy, respectively. The nuclear structural parts of
the cross section formulae are summarized by
\begin{eqnarray}
\lambda_{CC}&=&\frac{1}{4(2J_i+1)}
\left[\left|M(\tau_-)\right|^2
 +g_A^2\left|M(\sigma\tau_-)\right|^2\right]
=B({\rm F})+g_A^2B({\rm GT}^-)\\
\lambda_{NC}&=&\frac{1}{4(2J_i+1)}
 \left|f_AM(\sigma)+g_AM(\sigma\tau_z)\right|^2.
\end{eqnarray}
The $\lambda_{NC}$ values determined from the present $B({\rm GT}^-)$
($R^2=8.24$) and $B(\sigma)$ values are compared with the
estimations in the several works \cite{BERN92,HANN03,RAGH88} in
Table~\ref{tab:ccnc}.
The isovector and isoscalar components destructively contribute to
$\lambda_{NC}$ for the $5/2^-_2$ state although they constructively
contribute for the other states. The relative phases between the
isovector and isoscalar components were taken from the shell-model
predictions, which are in agreement with the results of the
$^{11}$B($p,\,p'$) analyses for the $1/2^-_1$,  $5/2^-_1$, and
$3/2^-_2$ states in $^{11}$B.

The value of $f_A/g_A$ is 
known to be 0.1-0.15 from the experiments \cite{ASHM89,ANTH00}, meaning
that the contribution from the axial isoscalar coupling to the NC
process increases the expected NC strength by 40\%. It is noted that the
isoscalar contributions were neglected in Refs.~\cite{HANN03,RAGH88}.

In Ref.~\cite{BERN92}, the $\lambda_{NC}$ value for the $1/2^-_1$ state
was determined from the experimental data without using any
particular nuclear model, but those for the other states were estimated
by combining the results from the experiments and from the shell-model
calculation. The estimated $\lambda_{NC}$ values given in
Ref.~\cite{BERN92} are consistent with the present experimental result.

\begin{table}
\caption{\label{tab:ccnc}
Transition strengths in the unit of $g_A^2$ for the neutrino induced
reactions via the neutral-current process in $^{11}$B . $f_A=0$ is
assumed in Refs.~\cite{HANN03} and \cite{RAGH88}.
}
\begin{ruledtabular}
\begin{tabular}{ccccc}
$J^\pi$ & 
\multicolumn{4}{c}{$\lambda_{NC}/g_A^2$}\\
& Present & 
Ref.~\cite{BERN92} & Ref.~\cite{HANN03} & Ref.~\cite{RAGH88}\\
\hline
$1/2^-_1$ & 
$0.101(8)\left[1+1.3(f_A/g_A)\right]^2$ &
$0.100\left[1+1.3(f_A/g_A)\right]^2$ &
0.068(13) & 0.089(6) \\
$5/2^-_1$ & 
$0.114(7)\left[1+0.8(f_A/g_A)\right]^2$ &
$0.114\left[1+1.0(f_A/g_A)\right]^2$ &
0.073(15) & 0.146(14) \\
$3/2^-_2$ & 
$0.120(8)\left[1+1.1(f_A/g_A)\right]^2$ &
$0.127\left[1+1.2(f_A/g_A)\right]^2$ &
0.088(18) & 0.146(6) \\
$5/2^-_2$ & 
$0.102(10)\left[1-0.7(f_A/g_A)\right]^2$ &
$-$ &
0.083(18) & $-$ \\
\end{tabular}
\end{ruledtabular}
\end{table}

%%%%%%%%%%%%%%%%%%%%%%%%%%%%%%%%%%%%%%%%%%%%%%%%%%%%%%%%%%%%%%%%%%%%%%%
\section{Summary}

We measured cross sections for the $^{11}$B($^3$He$,\,t$),
$^{11}$B($d,\,d'$), and $^{11}$B($p,\,p'$) reactions at forward
scattering angles including 0$^\circ$ to study the
isovector and isoscalar spin-flip M1 transition strengths. Analyzing
powers, induced polarizations, and depolarization parameters were
measured for the $^{11}$B($p,\,p'$) reaction. 

The measured $^{11}$B($^{3}$He$,\,t$) cross sections were compared with
the DWIA calculation using the POT wave functions. The
effective $^3$He-N interaction was obtained by fitting the cross
sections for the ground-state transition. The cross
sections for the transitions to the excited states in $^{11}$C were
decomposed into each ${\Delta}J^\pi$ component based on the DWIA
calculation, and the $B({\rm GT}^-)$ values were extracted from the
${\Delta}J^\pi=1^+$ contribution. The $B({\rm GT}^-)$ values are easily
converted to the isovector spin-flip M1 strengths $B(\sigma\tau_z)$ 
for the analogue transitions under the assumption of the isospin symmetry.

In the analysis of the $^{11}$B($d,\,d'$) reaction, we used the
$^{12}$C($d,\,d'$) reaction as a measure to obtain the angular
distribution of the cross section for each ${\Delta}J^\pi$ contribution.
After the angular distribution was obtained, the $^{11}$B($d,\,d'$) cross
section for each excited state was decomposed into the
${\Delta}J^\pi=0^+$, $1^+$, and $2^+$ components. Finally, the obtained
cross section for the ${\Delta}J^\pi=1^+$ component was converted to the
isoscalar spin-flip M1 strength $B(\sigma)$. 

The isovector and isoscalar transition strengths obtained in the
$^{11}$B($^3$He$,\,t$) and $^{11}$B($d,\,d'$) analyses were used for
performing the DWIA calculation to analyze the $^{11}$B($p,\,p'$)
reaction data. The
DWIA calculation reasonably explains the present $^{11}$B($p,\,p'$)
result except the cross section for the transition to the 8.92-MeV
states and supports the reliability of the present
$^{11}$B($^3$He$,\,t$) and
$^{11}$B($d,\,d'$) analyses. The DWIA calculation considerably
underestimates the cross section for the transition to the 8.92-MeV
state. This result implies that the 8.92-MeV
state has a nature totally different from the other three states at the
lower excited energies.

The obtained spin-flip M1 strengths were compared with the shell-model
calculations using the POT and SFO interactions. Although the
excitation energies for the excited states are reasonably in agreement
with the shell-model calculations, the transition strengths are
overestimated by 20-50\%. 
The transition strengths for the neutrino induced reactions were
estimated by using the isovector and isoscalar spin-flip M1 strengths.
The present results are quantitatively in agreement with the theoretical
estimation discussing the axial-isoscalar coupling in the neutrino
scattering process, and are useful in the future measurement of the
stellar neutrinos using the neutral- and charged-current reactions on
$^{11}$B.

\begin{acknowledgments}
The authors would like to thank Prof.~H.~Kamada, Dr.~Y.~Satou,
Prof.~Toshio~Suzuki, Prof.~T.~Otsuka, and Dr.~R.~Fujimoto for
valuable discussions of the theoretical calculations. They gratefully
acknowledge the effort of the RCNP cyclotron crew for
providing the stable and clean beam. This research was supported in part
by the Grant-in-Aid for Scientific Research No. 15740136 from the Japan
Ministry of Education, Sports, Culture, Science, and Technology.
\end{acknowledgments}

\bibliography{prc}

%%%%%%%%%%%%%%%%%%%%%%%%%%%%%%%%%%%%%%%%%%%%

%%%%%%%%%%%%%%%%%%%%%%%%%%%%%%%%%%%%%%%%%%%%%%%%%%%%%%%%%%%%%%%%%%%%%%%%%%

\end{document}